\documentclass{INTERSPEECH2023}
\usepackage{amsmath}
\usepackage{multirow}
\usepackage{amssymb}

\interspeechcameraready 

\title{CAM++: A Fast and Efficient Network for Speaker Verification Using Context-Aware Masking}

\name{Hui Wang, Siqi Zheng, Yafeng Chen, Luyao Cheng, Qian Chen}
\address{
	Speech Lab, Alibaba Group}
\email{\{tongmu.wh, zsq174630\}@alibaba-inc.com}

\begin{document}
\maketitle

\begin{abstract}
\renewcommand{\thefootnote}{\fnsymbol{footnote}}
Time delay neural network (TDNN) has been proven to be efficient for speaker verification. One of its successful variants, ECAPA-TDNN, achieved state-of-the-art performance at the cost of much higher computational complexity and slower inference speed. This makes it inadequate for scenarios with demanding inference rate and limited computational resources. We are thus interested in finding an architecture that can achieve the performance of ECAPA-TDNN and the efficiency of vanilla TDNN. 
In this paper, we propose an efficient network based on context-aware masking, namely CAM++, which uses densely connected time delay neural network (D-TDNN) as backbone and adopts a novel multi-granularity pooling to capture contextual information at different levels.
Extensive experiments on two public benchmarks, VoxCeleb and CN-Celeb, demonstrate that the proposed architecture outperforms other mainstream speaker verification systems with lower computational cost and faster inference speed. \footnote[2]{The source code is available at \url{https://github.com/alibaba-damo-academy/3D-Speaker}}
	
\end{abstract}

\section{Introduction}
Speaker verification (SV) is the task of automatically verifying whether an utterance is pronounced by a hypothesized speaker based on the voice characteristic \cite{journals-BaiZ21}.
Typically, a speaker verification system consists of two main components -  
an embedding extractor which transforms an utterance of random length into a fixed-dimensional speaker embedding, and a back-end model that calculates the similarity score between the embeddings \cite{{xvector,zheng2019autoencoder}}.

Over past few years, speaker verification systems based on deep learning methods~\cite{xvector,ecapa,DTDNN,CAM,pacnet} have achieved remarkable improvements. 
One of the most popular systems is x-vector, which adopts time delay neural network (TDNN) as backbone. TDNN takes one-dimensional convolution along the time axis to capture local temporal context information. 
Following the successful application of x-vector, several modifications are proposed to enhance robustness of the networks. 
ECAPA-TDNN~\cite{ecapa} unifies one-dimensional Res2Block with squeeze-excitation~\cite{se} and expands the temporal context of each layer, achieving significant improvement. 
At the same time, the topology of x-vector is improved by incorporating elements of ResNet~\cite{resnet} which uses a two-dimensional convolutional neural network (CNN) with convolutions in both time and frequency axes. Equiped with residual connection, ResNet-based systems~\cite{rvector,df-resnet} have achieved outstanding results. 
However, these networks tend to require a large number of parameters and computations to achieve optimal performance. 
In real-world applications, accuracy and efficiency are equally important. It is of sufficient interest and challenge to find a speaker embedding extracting network that simultaneously improves the performance, computation complexity, and inference speed.

Recently, \cite{DTDNN} proposes a TDNN-based architecture, called densely connected time delay neural network (D-TDNN), by adopting bottleneck layers and dense connectivity. It obtains better accuracy with fewer parameters compared to vanilla TDNN. 
Later, in~\cite{CAM}, a context-aware masking (CAM) module is proposed to make the D-TDNN focus on the speaker of interest and ``blur" unrelated noise, while requiring only a little computation cost. 
Despite of significant improvements on accuracy, there still exists a large performance gap compared to other state-of-the-art speaker models~\cite{ecapa}. 

\begin{figure}[t]
	\centering
	\includegraphics[width=0.9\linewidth]{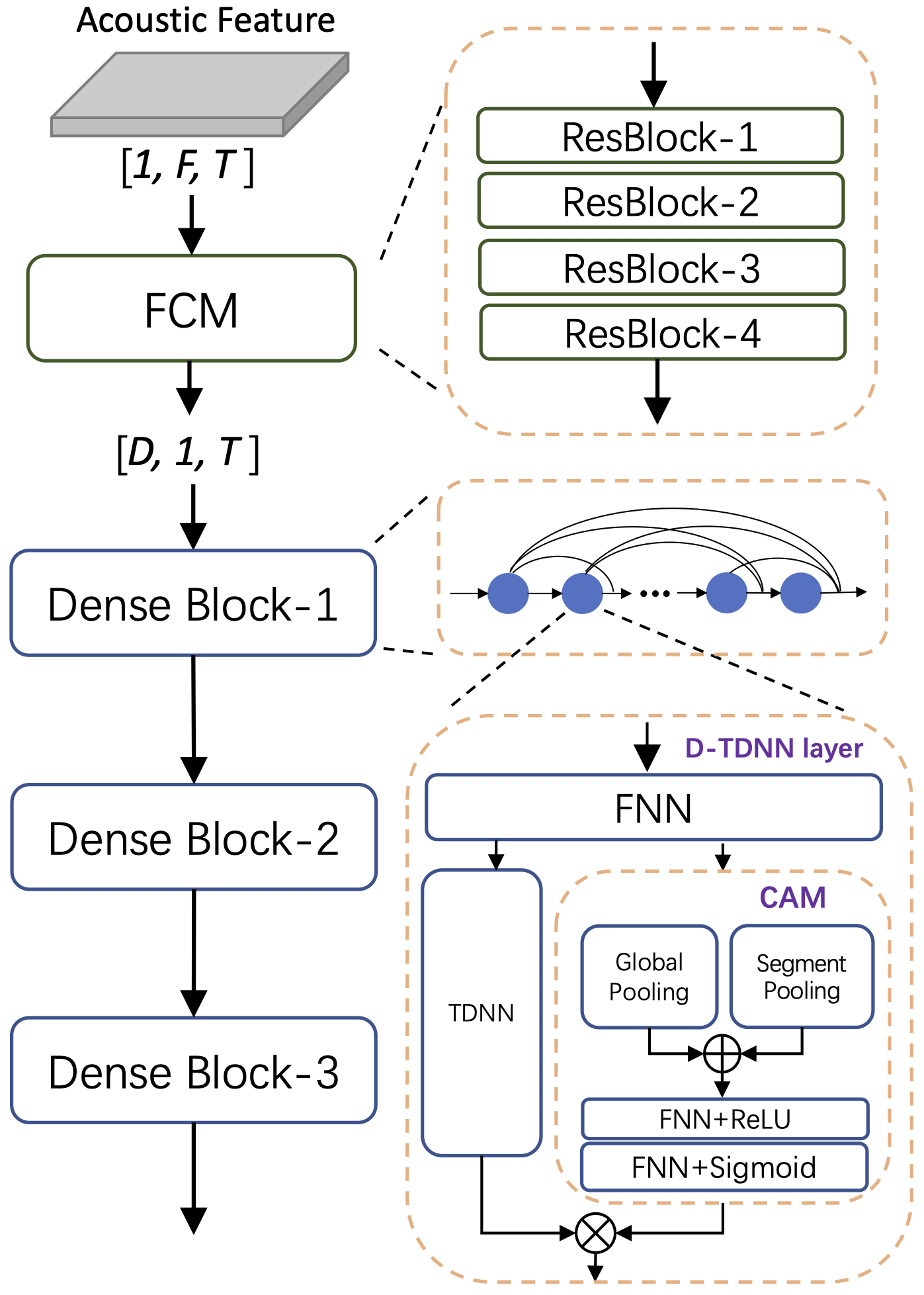}
	\caption{Overview of the proposed CAM++ architecture. It comprises convolution modules as the front-end and D-TDNN as the backbone. An improved context-aware making is built into each D-TDNN layer, which includes multi-granularity pooling to capture speaker characteristics. }
	\label{fig:structure}
\end{figure}

In this paper, we propose CAM++, an efficient and accurate network for speaker embedding learning that utilizes D-TDNN as a backbone, as shown in Figure~\ref{fig:structure}. We have adopted multiple methodologies to enhance the CAM module and D-TDNN architecture.
Firstly, we design a lighter CAM module and insert it into each D-TDNN layer to place more focus on the speaker characteristics of interest. Multi-granularity pooling is an essential component of the CAM module, built to capture contextual information at both global and segment levels. The previous study in~\cite{ponet} showed that multi-granularity pooling achieves comparable performance with much higher efficiency, when compared to a transformer structure. 
Secondly, we adopt a narrower network with fewer filters in each D-TDNN layer, significantly increasing the network depth compared to vanilla D-TDNN~\cite{DTDNN}. This is motivated by ~\cite{df-resnet}, which observed that deeper layers can bring more improvements than wider channels for speaker verification. 
Finally, we incorporate a two-dimensional convolution module as a front-end to enhance the D-TDNN network's ability to be invariant to frequency shifts in the input features. A hybrid architecture of TDNN and CNN has been shown to yield further improvements~\cite{cnn-ecapa,mfa-tdnn}.
We evaluate the proposed architecture on two public benchmarks, VoxCeleb~\cite{voxceleb} and CN-Celeb~\cite{cnceleb1,cnceleb2}. The results show that our method obtains 0.73\% and 6.78\% EER in VoxCeleb-O and CN-Celeb test sets. Furthermore, our architecture has lower computation complexity and faster inference speed than popular ECAPA-TDNN and ResNet34 systems.

\section{System description}
\subsection{Overview}
\label{sec:overview}
The overall framework of the proposed CAM++ architecture is illustrated in Figure~\ref{fig:structure}. The architecture mainly consists of two components: the front-end convolution module (FCM) and the D-TDNN backbone. The FCM consists of multiple blocks of two-dimensional convolution with residual connections, which encode acoustic features in the time-frequency domain to exploit high-resolution time-frequency details. The resulting feature map is subsequently flattened along the channel and frequency dimensions and used as input for the D-TDNN. The D-TDNN backbone comprises three blocks, each containing a sequence of D-TDNN layers.
In each D-TDNN layer, we build an improved CAM module that assigns different attention weights to the output feature of the inner TDNN layer. The multi-granularity pooling incorporates global average pooling and segment average pooling to effectively aggregate contextual information across different levels. With dense connections, the masked output is concatenated with all preceding layers and serves as the input for the next layer.

\subsection{D-TDNN backbone}
\label{sec:D-TDNN-net}
TDNN uses a dilated one-dimensional convolution structure along the time axis as its backbone, which was first adopted by x-vector~\cite{xvector}. Due to its success, TDNN has been widely used in speaker verification tasks. 
An improved version, D-TDNN, was recently proposed in~\cite{DTDNN} as an efficient TDNN-based speaker embedding model. Similar to DenseNet~\cite{densenet}, it adopts dense connectivity, which involves direct connections among all layers in a feed-forward manner. 
D-TDNN is parameter-efficient and achieves better results while requiring fewer parameters than vanilla TDNN. Hence, we adopt D-TDNN as the backbone of our network.

Specifically, the basic unit of D-TDNN consists of a feed-forward neural
network (FNN) and a TDNN layer. A direct connection is applied between the input of two consecutive D-TDNN layers. The formulation of the $l$-th D-TDNN layer is:
\begin{align}
	\boldsymbol{S}^l = \mathcal{H}_l([\boldsymbol{S}^0, \boldsymbol{S}^1, \cdots, \boldsymbol{S}^{l-1}])
\end{align}
where $\boldsymbol{S}^0$ is the input of the D-TDNN block, $\boldsymbol{S}^l$ is the output of the the $l$-th D-TDNN layer, $\mathcal{H}_l$ denotes the non-linear transformation of the $l$-th D-TDNN layer. 

Although D-TDNN has demonstrated remarkable improvement in comparison to vanilla TDNN, there remains a considerable gap between it and state-of-the-art speaker embedding models like ECAPA-TDNN and ResNet34. We redesign the D-TDNN to further push its limits and achieve better results. 
In~\cite{df-resnet}, it is revealed that depth of the network plays a critical role in the performance of speaker verification, and increasing the depth of the speaker embedding model tends to yield more improvement than widening it.
Hence, we significantly increase the depth of the D-TDNN network while reducing the channel size of filters in each layer to control the network's complexity.
Specifically, the vanilla D-TDNN has two blocks, each containing 6 and 12 D-TDNN layers, respectively. We add an additional block at the end and expand the number of layers per block to 12, 24 and 16. To reduce the network's complexity, we adopt narrower D-TDNN layers in each block, that is, reducing the original growth rate $k$ from 64 to 32. Additionally, we adopted an input TDNN layer with 1/2 subsampling rate before the D-TDNN backbone to accelerate computation. In Section~\ref{overall-results}, the experimental results will indicate that these effective modifications significantly improve the performance of speaker verification.

\subsection{Context-aware masking}
\label{sec:pooling-mask}
Attention mechanism has been widely adopted in speaker verification.  
Squeeze-excitation (SE)~\cite{se} squeezes global spatial information into a channel descriptor to model channel interdependencies and recalibrate filter responses. 
Meanwhile, soft self-attention is utilized to calculate the weighted statistics for the improvement of temporal pooling techniques~\cite{attentive-pooling,multihead-pooling1,multihead-pooling2}.

An attention-based context-aware masking (CAM) module was recently proposed in~\cite{CAM} to focus on the speaker of interest and blur unrelated noise, resulting a significant improvement in the performance of D-TDNN.
CAM performs feature map masking using an auxiliary utterance-level embedding obtained from global statistic pooling. 
However, in~\cite{CAM}, CAM is only applied at the transition layer after each D-TDNN block, and a limited number of CAM modules may be insufficient for extracting critical information effectively.
To address this, we propose a lighter CAM and insert it into each D-TDNN layer to capture more speaker characteristic of interest.  

As shown in Figure~\ref{fig:structure}, we denote the output hidden feature from the head FNN in the D-TDNN block as $\boldsymbol{X}$. Firstly, $\boldsymbol{X}$ is input into the TDNN layer to extract local temporal feature $\boldsymbol{F}$:
\begin{align}
	\boldsymbol{F} = \mathcal{F}(\boldsymbol{X})
\end{align}
where $\mathcal{F}(\cdot)$ denotes the transformation of the TDNN layer. $\mathcal{F}(\cdot)$ only focuses on local receptive field and $\boldsymbol{F}$ may be suboptimal. Therefore, 
a ratio mask $\boldsymbol{M}$ is predicted based on an extracted contextual embedding $\boldsymbol{e}$, and is expected to contain both speaker of interest and noise characteristic. 
\begin{align}
	\label{equation:mask}
	\boldsymbol{M}_{*t} &= \sigma(\boldsymbol{W}_2\delta(\boldsymbol{W}_1\boldsymbol{e}+\boldsymbol{b}_1)+\boldsymbol{b}_2)
\end{align}
where $\sigma(\cdot)$ denotes the Sigmoid function, $\delta(\cdot)$ denotes the ReLU function, and $\boldsymbol{M}_{*t}$ denotes the $t$-th frame of $\boldsymbol{M}$. 

In~\cite{CAM}, a global statistic pooling is used to generate the contextual embedding $\boldsymbol{e}$. 
It is known that speech signals have typical hierarchical structure and exhibit dynamic changes in characteristic between different subsegments. A unique speaking manner of the target speaker may exist within a certain segment. 
Simply using a single embedding from global pooling may result in loss of precise local contextual information, leading to a suboptimal masking. 
Therefore, it is beneficial to extend the single global pooling to multi-granularity pooling. This enables the network to capture more contextual information at different levels, generating a more accurate mask. Specifically, a global average pooling is used to extract contextual information at global level:
\begin{align}
	\boldsymbol{e}_g &= \frac{1}{T}\sum_{t=1}^{T}\boldsymbol{X}_{*t}
\end{align}
Simultaneously, a segment average pooling is used to extract contextual information at segment level:
\begin{equation}
	\boldsymbol{e}_s^k = \frac{1}{s_{k+1}-s_k}\sum_{t=s_k}^{s_{k+1}-1}\boldsymbol{X}_{*t}
\end{equation}
Where $s_{k}$ is the starting frames of $k$-th segment of feature $\boldsymbol{X}$. In the experiments, we segment the frame-level feature $\boldsymbol{X}$ into consecutive fixed-length 100-frame segments and apply segment average pooling to each.

Subsequently, contextual embeddings of different level , $\boldsymbol{e}_g$ and $\boldsymbol{e}_s$, are aggregated to predict the context-aware mask $\boldsymbol{M}$. The Equation~\ref{equation:mask} can be rewrote to:
\begin{align}
	\label{equation:mask_modified}
	\boldsymbol{M}_{*t}^k =& \sigma(\boldsymbol{W}_2\delta(\boldsymbol{W}_1(\boldsymbol{e}_g+\boldsymbol{e}_s^k)+\boldsymbol{b}_1)+\boldsymbol{b}_2), \nonumber \\
	& s_k \leqslant t < s_{k+1}
\end{align}
Finally, predicted $\boldsymbol{M}$ is used to calibrate the representation and produce the refined representation $\tilde{\boldsymbol{F}}$.
\begin{align}
	\tilde{\boldsymbol{F}} &= \mathcal{F}(\boldsymbol{X}) \odot \boldsymbol{M}
\end{align}
Where $\odot$ denotes the element-wise multiplication.
Equation~\ref{equation:mask_modified} has a simpler form and fewer trainable parameters compared to~\cite{CAM}. We insert this efficient context-aware masking into each D-TDNN layer to enhance the representational power of basic layers throughout the network. 

\subsection{Front-end convolution module}
TDNN-based networks perform one-dimension convolution along the time axis, using kernels that cover the complete frequency range of the input features.
It is more difficult to capture speaker characteristics occurring at certain local frequency regions compared to two-dimensional convolutional network~\cite{cnn-ecapa}. 
Generally, plenty of filters are required to model the complex details in the full frequency region.
For examples, ECAPA-TDNN has a maximum of 1024 channels in the convolutional layers to achieve optimal performance. 
In Section~\ref{sec:D-TDNN-net}, we use narrower layers in each D-TDNN block to control the size of parameters. This may result in a reduced ability to find the specific frequency pattern in some local regions. 
It is necessary to enhance the robustness of D-TDNN to small and reasonable shifts in the time-frequency domain and compensate for realistic intra-speaker pronunciation variability. Motivated by~\cite{cnn-ecapa,mfa-tdnn}, we equip the D-TDNN network with a two-dimensional front-end convolution module (FCM). 
Inspired by the success of ResNet-based architectures in speaker verification, we decide to incorporate 4 residual blocks in the FCM stem, as illustrated in Figure~\ref{fig:structure}. 
The number of channels is set to 32 for all residual blocks. We use a stride of 2 in the frequency dimension in the last three blocks, resulting in an 8x downsampling in the frequency domain. 
The output feature map of FCM is subsequently flattened along the channel and frequency dimension and used as input for the D-TDNN backbone.

\section{Experiments}

\subsection{Dataset}
We conduct experiments on two public speaker verification benchmarks, VoxCeleb~\cite{voxceleb} and CN-Celeb~\cite{cnceleb1,cnceleb2}, to  evaluate the effectiveness of the proposed methods. For VoxCeleb, we use the development set of VoxCeleb2 for training, which contains 5,994 speakers. The evaluation set is constructed from three cleaned version test trials, VoxCeleb1-O, VoxCeleb1-E and VoxCeleb1-H. The last two tasks have more trial pairs. For CN-Celeb, the development sets of CN-Celeb1 and CN-Celeb2 are used for training, which contain 2785 speakers. In the data preprocessing of the training data, we concatenate short utterances to ensure that they are no less than 6s. There exists multiple utterances for each enrollment speaker in CN-Celeb test set. We choose to average all the embeddings which belong to the same enrollment speaker to get final speaker embedding for evaluation.
\subsection{Experimental setup}
\label{sec:setup}
For all experiments, we use 80-dimensional Fbank features extracted over a 25 ms long window for every 10 ms as input. 
We apply speed perturbation augmentation by randomly sampling a ratio from $\{0.9, 1.0, 1.1\}$. The processed audio is considered to be from a new speaker~\cite{sresystem}.  In addition, two popular data augmentations are adopted during training, simulating reverberation using the RIR dataset~\cite{rirs}, adding noise using the MUSAN dataset~\cite{musan}. 

Angular additive margin softmax (AAM-Softmax) loss~\cite{AAM} is used for all experiments. The margin and scaling factors of AAM-Softmax loss are set to 0.2 and 32 respectively.
During training, we adopt stochastic gradient descent (SGD) optimizer with a cosine annealing scheduler and a linear warm-up scheduler, where the learning rate is varied between 0.1 and 1e-4. The momentum is 0.9, and the weight decay is 1e-4. 3s-long samples are randomly cropped from each audio to construct the training minibatches. 

We use cosine similarity scoring for evaluation, without applying score normalization in the back-end. We adopt two commonly used metrics in speaker verification tasks, equal error rate (EER) and the minimum detection cost function (MinDCF) with 0.01 target probability. 

%
\begin{table*}[]
	
	\caption{Performance comparison of different network architectures on the VoxCeleb1 and CN-Celeb test sets. Data augmentation strategy and experimental setup are kept consistent throughout all experiments.}
	\label{tab:results}
	\centering
	\begin{tabular}{cccccc}
		\bottomrule
		\hline
		\multirow{2}{*}{\textbf{Architecture}} & \multirow{2}{*}{\textbf{Params(M)}} & \textbf{VoxCeleb1-O} & \textbf{VoxCeleb1-E} & \textbf{VoxCeleb1-H} & \textbf{CN-Celeb Test} \\
		&  & \textbf{EER(\%)/MinDCF}& \textbf{EER(\%)/MinDCF}& \textbf{EER(\%)/MinDCF}& \textbf{EER(\%)/MinDCF}\\ \hline
		TDNN &4.62        &2.31/0.3223      & 2.37/0.2732    & 4.25/0.3931       &9.86/0.6199         \\
		ECAPA-TDNN&14.66     & 0.89/0.0921          & 1.07/0.1185         & 1.98/0.1956          & 7.45/0.4127     \\ \hline
        ResNet34  & 6.70                  & 0.97/0.0877          & 1.03/0.1133         &1.88/0.1778           & 6.97/0.3859         \\ \hline 
		D-TDNN    & 2.85                 & 1.55/0.1656          & 1.63/0.1748         & 2.86/0.2571          & 8.41/0.4683         \\
		D-TDNN-L & 6.40                  & 1.19/0.1179          & 1.21/0.1287         & 2.22/0.2047          & 7.82/0.4336         \\ \hline
		\textbf{CAM++}  &7.18                 & \textbf{0.73/0.0911}          & \textbf{0.89/0.0995}         & \textbf{1.76/0.1729}          & \textbf{6.78/0.3830}         \\ 
		-w/o Masking& 6.64      & 0.93/0.1022          & 1.03/0.1144         & 1.86/0.1762          & 7.16/0.3947         \\
		-w/o FCM&6.94           & 0.98/0.1127          & 1.01/0.1175        & 2.03/0.2006          & 7.17/0.4011         \\
		\bottomrule
	\end{tabular}
\end{table*}
%

%
%


%
\begin{table}[]
	\caption{Performance comparison of multiple key components of CAM++. GP represents masking with only global pooling and SP denotes segment pooling.}
	\label{tab:compare}
	\centering
	\begin{tabular}{cccccccccc}
		\bottomrule
		\hline
		\multirow{2}{*}{\textbf{Method}} &  \multirow{2}{*}{\textbf{Params(M)}} &
		\multicolumn{2}{c}{\textbf{CN-Celeb Test}} \\ \cline{3-4}
		&&\textbf{EER(\%)}&\textbf{MinDCF}\\ \hline
		D-TDNN & 2.85 & 8.41 & 0.4683\\ \hline
		CAM~\cite{CAM} & 4.10 & 7.80 & 0.4431 \\ 
		GP& 3.07 & 7.78 & 0.4321\\ 
		\textbf{GP+SP}& 3.07 & \textbf{7.59} & \textbf{0.4209}\\
		\bottomrule
	\end{tabular}
\end{table}

\subsection{Results on VoxCeleb and CN-Celeb}
\label{overall-results}
The performance overview of all methods is presented in Table~\ref{tab:results}. For fair comparison, we re-implement several baseline models under the same experimental setup described in Section~\ref{sec:setup}, including TDNN~\cite{xvector}, D-TDNN~\cite{DTDNN}, ECAPA-TDNN ~\cite{ecapa} and ResNet34~\cite{rvector}. The ResNet34 model contains four residual blocks with different channel sizes, [64, 128, 256, 512], in each block. The ECAPA-TDNN model with 1024 channels is built according to~\cite{ecapa}.

It can be found in Table~\ref{tab:results} that, as an improved variant of TDNN, ECAPA-TDNN achieves impressive improvement in EER and MinDCF but requires large amounts of parameters. 
Using dense connection, D-TDNN outperforms TDNN with fewer parameters.
Compared to the standard D-TDNN, it can be found that deeper D-TDNN-L proposed in Section~\ref{sec:D-TDNN-net} achieves significant performance improvement, thanks to increased parameters and effective modifications.
However, there is still a large performance gap compared to ECAPA-TDNN or ResNet34. 
When we equip the D-TDNN-L backbone with CAM with multi-granularity pooling and FCM, CAM++ consistently performs better than the ECAPA-TDNN and ResNet34 baselines.
In particular, CAM++ has relative 51\% fewer parameters and 18\% lower EER than ECAPA-TDNN in VoxCeleb-O. 

Next, we remove individual components to explore the contribution of each to the performance improvements.
It can be observed that CAM with multi-granularity pooling improves the EER in VoxCeleb-O and CN-Celeb test sets by 21\% and 5\%, respectively. This confirms the benefit of aggregating contextual vectors at different levels to perform attention masking.
Removing FCM leads to a obvious increase in EER and MinDCF in all test sets. 
This phenomenon indicates that stronger speaker embeddings can be obtained from a hybrid of two-dimensional convolution and TDNN-based network. 

\subsection{Impacts of multi-granularity pooling}
We further evaluate the effectiveness of the improved CAM with multi-granularity pooling. Additional experimental results on the CN-Celeb test set are presented in Table~\ref{tab:compare}.
We use D-TDNN~\cite{DTDNN} as the baseline. We re-implement the CAM proposed in ~\cite{CAM} on CN-Celeb, and find it decrease the EER by 7\% relatively but with a 44\% increase in parameters. 
Next, We apply the improved CAM proposed in Section~\ref{sec:pooling-mask} to D-TDNN only with global average pooling (GP), which results in similar improvement in EER with only an 8\% increase in parameters, demonstrating better parameters efficiency.
We then apply segment average pooling (SP) and fuse it with GP, observing performance gains without introducing additional parameters. These results indicate the importance of local segment contextual information in performing more accurate masking.

\subsection{Complexity analysis}
In this section, we compare the complexity of ECAPA-TDNN, ResNet34 and CAM++ models, including the number of parameters, floating-point operations (FLOPs) and real-time factor (RTF), as shown in Table~\ref{tab:complexity}. RTF was evaluated on the CPU device under single-thread condition. When comparing CAM++ with ResNet34, CAM++ has slightly more parameters but significant fewer FLOPs. At the same time, CAM++ has half the parameters and FLOPs of ECAPA-TDNN. It is worth noting that CAM++ achieves more than twice the inference speed of both ResNet34 and ECAPA-TDNN.  Although ResNet34 and ECAPA-TDNN have a similar RTF, they have different FLOPs. This is likely due to increased memory access resulting from higher parameter data dependencies, which leads to increased computation time.

\begin{table}[]
	\caption{The number of parameters, floating-point operations (FLOPs) and real-time factor (RTF) of different models. RTF was evaluated on CPU under single-thread condition.}
	\label{tab:complexity}
	\centering
	\begin{tabular}{cccc}
		\bottomrule
		\hline
		\textbf{Model} & \textbf{Params(M)} & \textbf{FLOPs(G)} & \textbf{RTF} \\ \hline 
		ECAPA-TDNN    & 14.66 & 3.96 & 0.033 \\ 
        ResNet34    & 6.70 & 6.84 & 0.032 \\
		\textbf{CAM++}    & 7.18 & \textbf{1.72} & \textbf{0.013} \\
		\bottomrule
	\end{tabular}
\end{table}

\section{Conclusion}
This paper proposed CAM++, an efficient speaker embedding model for speaker verification. Our novel context-aware masking method aimed to focus on the speaker of interest and improved the quality of features, while multi-granularity pooling fused different levels of contextual information to generate accurate attention weights. We conducted comprehensive experiments on two public benchmarks, VoxCeleb and CN-Celeb. The results demonstrated that CAM++ achieved superior performance with lower computational complexity and faster inference speed than popular ECAPA-TDNN and ResNet34 systems. 

\bibliographystyle{IEEEtran}
\bibliography{mybib}

\begin{thebibliography}{10}
\providecommand{\url}[1]{#1}
\csname url@samestyle\endcsname
\providecommand{\newblock}{\relax}
\providecommand{\bibinfo}[2]{#2}
\providecommand{\BIBentrySTDinterwordspacing}{\spaceskip=0pt\relax}
\providecommand{\BIBentryALTinterwordstretchfactor}{4}
\providecommand{\BIBentryALTinterwordspacing}{\spaceskip=\fontdimen2\font plus
\BIBentryALTinterwordstretchfactor\fontdimen3\font minus
  \fontdimen4\font\relax}
\providecommand{\BIBforeignlanguage}[2]{{%
\expandafter\ifx\csname l@#1\endcsname\relax
\typeout{** WARNING: IEEEtran.bst: No hyphenation pattern has been}%
\typeout{** loaded for the language `#1'. Using the pattern for}%
\typeout{** the default language instead.}%
\else
\language=\csname l@#1\endcsname
\fi
#2}}
\providecommand{\BIBdecl}{\relax}
\BIBdecl

\bibitem{journals-BaiZ21}
Z.~Bai and X.~Zhang, ``Speaker recognition based on deep learning: An
  overview,'' \emph{Neural Networks}, vol. 140, pp. 65--99, 2021.

\bibitem{xvector}
D.~Snyder, D.~Garcia-Romero, G.~Sell, D.~Povey, and S.~Khudanpur, ``X-vectors:
  Robust dnn embeddings for speaker recognition,'' in \emph{IEEE International
  Conference on Acoustics, Speech and Signal Processing (ICASSP)}, 2018, pp.
  5329--5333.

\bibitem{zheng2019autoencoder}
S.~Zheng, G.~Liu, H.~Suo, and Y.~Lei, ``Autoencoder-based semi-supervised
  curriculum learning for out-of-domain speaker verification,'' in \emph{Annual
  Conference of the International Speech Communication Association
  (INTERSPEECH)}, 2019, pp. 4360--4364.

\bibitem{ecapa}
B.~Desplanques, J.~Thienpondt, and K.~Demuynck, ``{ECAPA-TDNN}: Emphasized
  channel attention, propagation and aggregation in tdnn based speaker
  verification,'' in \emph{Annual Conference of the International Speech
  Communication Association (INTERSPEECH)}, 2020, pp. 3830--3834.

\bibitem{DTDNN}
Y.-Q. Yu and W.-J. Li, ``Densely connected time delay neural network for
  speaker verification,'' in \emph{Annual Conference of the International
  Speech Communication Association (INTERSPEECH)}, 2020, pp. 921--925.

\bibitem{CAM}
Y.-Q. Yu, S.~Zheng, H.~Suo, Y.~Lei, and W.-J. Li, ``Cam: Context-aware masking
  for robust speaker verification,'' in \emph{IEEE International Conference on
  Acoustics, Speech and Signal Processing (ICASSP)}, 2021, pp. 6703--6707.

\bibitem{pacnet}
S.~Zheng, Y.~Lei, and H.~Suo, ``Phonetically-aware coupled network for short
  duration text-independent speaker verification,'' in \emph{Interspeech 2020,
  21st Annual Conference of the International Speech Communication
  Association}.\hskip 1em plus 0.5em minus 0.4em\relax {ISCA}, 2020, pp.
  926--930.

\bibitem{se}
J.~Hu, L.~Shen, and G.~Sun, ``Squeeze-and-excitation networks,'' in \emph{IEEE
  Conference on Computer Vision and Pattern Recognition (CVPR)}, 2018, pp.
  7132--7141.

\bibitem{resnet}
K.~He, X.~Zhang, S.~Ren, and J.~Sun, ``Deep residual learning for image
  recognition,'' in \emph{IEEE Conference on Computer Vision and Pattern
  Recognition (CVPR)}, 2016, pp. 770--778.

\bibitem{rvector}
H.~Zeinali, S.~Wang, A.~Silnova, P.~Matejka, and O.~Plchot, ``But system
  description to voxceleb speaker recognition challenge 2019,'' 2019,
  arXiv:1910.12592.

\bibitem{df-resnet}
B.~Liu, Z.~Chen, S.~Wang, H.~Wang, B.~Han, and Y.~Qian, ``Df-resnet: Boosting
  speaker verification performance withdepth-first design,'' in \emph{Annual
  Conference of the International Speech Communication Association
  (INTERSPEECH)}, 2022, pp. 296--300.

\bibitem{ponet}
C.~Tan, Q.~Chen, W.~Wang, Q.~Zhang, S.~Zheng, and Z.~Ling, ``Ponet: Pooling
  network for efficient token mixing in long sequences,'' in \emph{The Tenth
  International Conference on Learning Representations, {ICLR} 2022, Virtual
  Event, April 25-29, 2022}.

\bibitem{cnn-ecapa}
J.~Thienpondt, B.~Desplanques, and K.~Demuynck, ``Integrating frequency
  translational invariance in tdnns and frequency positional information in 2d
  resnets to enhance speaker verification,'' in \emph{Annual Conference of the
  International Speech Communication Association (INTERSPEECH)}, 2021, pp.
  2302--2306.

\bibitem{mfa-tdnn}
T.~Liu, R.~K. Das, K.~Aik~Lee, and H.~Li, ``{MFA}: {TDNN} with multi-scale
  frequency-channel attention for text-independent speaker verification with
  short utterances,'' in \emph{IEEE International Conference on Acoustics,
  Speech and Signal Processing (ICASSP)}, 2022, pp. 7517--7521.

\bibitem{voxceleb}
A.~Nagrani, J.~S. Chung, W.~Xie, , and A.~Zisserman, ``Voxceleb: Large-scale
  speaker verification in the wild,'' \emph{Computer Speech and Language},
  vol.~60, 2020.

\bibitem{cnceleb1}
Y.~Fan, J.~Kang, L.~Li, K.~Li, H.~Chen, S.~Cheng, P.~Zhang, Z.~Zhou, Y.~Cai,
  and D.~Wang, ``{CN-Celeb}: a challenging chinese speaker recognition
  dataset,'' in \emph{IEEE International Conference on Acoustics, Speech and
  Signal Processing (ICASSP)}.\hskip 1em plus 0.5em minus 0.4em\relax IEEE,
  2020, pp. 7604--7608.

\bibitem{cnceleb2}
L.~Li, R.~Liu, J.~Kang, Y.~Fan, H.~Cui, Y.~Cai, R.~Vipperla, T.~F. Zheng, and
  D.~Wang, ``{CN-Celeb}: multi-genre speaker recognition,'' \emph{Speech
  Communication}, 2022.

\bibitem{densenet}
G.~Huang, Z.~Liu, L.~van~der Maaten, and K.~Q. Weinberger, ``Densely connected
  convolutional networks,'' in \emph{IEEE Conference on Computer Vision and
  Pattern Recognition (CVPR)}, 2017, p. 2261–2269.

\bibitem{attentive-pooling}
K.~Okabe, T.~Koshinaka, and K.~Shinoda, ``Attentive statistics pooling for deep
  speaker embedding,'' in \emph{Annual Conference of the International Speech
  Communication Association (INTERSPEECH)}, 2018, pp. 2252--2256.

\bibitem{multihead-pooling1}
Y.~Zhu, T.~Ko, D.~Snyder, B.~Mak, and D.~Povey, ``Self-attentive speaker
  embeddings for text independent speaker verification,'' in \emph{Annual
  Conference of the International Speech Communication Association
  (INTERSPEECH)}, 2018, pp. 3573--3577.

\bibitem{multihead-pooling2}
M.~Indiay, P.~Safariy, and J.~Hernando, ``Self multi-head attention for speaker
  recognition,'' in \emph{Annual Conference of the International Speech
  Communication Association (INTERSPEECH)}, 2019, pp. 4305--4309.

\bibitem{sresystem}
Z.~Chen, B.~Han, X.~Xiang, H.~Huang, B.~Liu, and Y.~Qian, ``Build a sre
  challenge system: Lessons from voxsrc 2022 and cnsrc 2022,'' 2022,
  arXiv:2211.00815v1.

\bibitem{rirs}
T.~Ko, V.~Peddinti, D.~Povey, M.~L. Seltzer, and S.~Khudanpur, ``A study on
  data augmentation of reverberant speech for robust speech recognition,'' in
  \emph{IEEE International Conference on Acoustics, Speech and Signal
  Processing (ICASSP)}, 2017, pp. 5220--5224.

\bibitem{musan}
D.~Snyder, G.~Chen, and D.~Povey, ``{MUSAN}: {A} {M}usic, {S}peech, and {N}oise
  {C}orpus,'' 2015, arXiv:1510.08484v1.

\bibitem{AAM}
J.~Deng, J.~Guo, N.~Xue, and S.~Zafeiriou, ``Arc-face: Additive angular margin
  loss for deep face recognition,'' in \emph{IEEE Conference on Computer Vision
  and Pattern Recognition (CVPR)}, 2019, pp. 4690--4699.

\end{thebibliography}
	
\end{document}